\documentclass{llncs}
\usepackage{amssymb}
\usepackage{graphicx}
\usepackage{booktabs}
\usepackage{dcolumn}
\usepackage{amsmath}

\usepackage{color}
\definecolor{Blue}{rgb}{0,0,1}

\definecolor{Orange}{rgb}{1,0.5,0}

\definecolor{Green}{rgb}{0,1,0}

\usepackage{url}
\urldef{\mailsa}\path|{alfred.hofmann, ursula.barth, ingrid.haas, frank.holzwarth,|
	\urldef{\mailsb}\path|anna.kramer, leonie.kunz, christine.reiss, nicole.sator,|
	\urldef{\mailsc}\path|erika.siebert-cole, peter.strasser, lncs}@springer.com|

\begin{document}
	
	\mainmatter  
	
\title{Mobile Communication Signatures of Unemployment}
	
	\titlerunning{Mobile Communication Signatures of Unemployment}
	
	%
	%
	\author{Abdullah Almaatouq\inst{1}
		\and Francisco Prieto-Castrillo~\inst{1,2}
		\and Alex Pentland\inst{1}
		}
	\authorrunning{Almaatouq et al.}
	\institute{Media Lab, Massachusetts Institute of Technology, Cambridge, MA, USA 
		\and Harvard T.H. Chan School of Public Health, Boston, MA 02115\\
		\email{\{amaatouq,pentland\}@mit.edu}\\
		\email{fprieto@hsph.harvard.edu}
	}
	
	%
	%
	
	\toctitle{Lecture Notes in Computer Science}
	\tocauthor{Authors' Instructions}
	\maketitle

\begin{abstract}
The mapping of populations socio-economic well-being is highly constrained by the logistics of censuses and surveys. Consequently, spatially detailed changes across scales of days, weeks, or months, or even year to year, are difficult to assess; thus the speed of which policies can be designed and evaluated is limited.
However, recent studies have shown the value of mobile phone data as an enabling methodology for demographic modeling and measurement. In this work, we investigate whether indicators extracted from mobile phone usage can reveal information about the socio-economical status of microregions such as districts (i.e., average spatial resolution $< 2.7km$). For this we examine anonymized mobile phone metadata combined with beneficiaries records from unemployment benefit program. We find that aggregated activity, social, and mobility patterns strongly correlate with unemployment. Furthermore, we construct a simple model to produce accurate reconstruction of district level unemployment from their mobile communication patterns alone. 
Our results suggest that reliable and cost-effective economical indicators could be built based on passively collected and anonymized mobile phone data. With similar data being collected every day by telecommunication services across the world, survey-based methods of measuring community socioeconomic status could potentially be augmented or replaced by such passive sensing methods in the future.
\end{abstract}

\section{Introduction}
As is well known, a major challenge in the development space is the lack of access to reliable and timely socio-economic data. Much of our understanding of the factors that affect the economical development of cities has been traditionally obtained through complex and costly surveys, with an update rate ranging from months to decades, which limits the scope of the studies and potentially bias the data~\cite{henrich2001search}. In addition, as participation rates in unemployment surveys drop, serious questions regarding the declining accuracy and increased bias in unemployment numbers have been raised~\cite{krueger2014evolution}. However, recent wide-spread adoption of electronic and pervasive technologies (e.g., mobile penetration rate of 100\% in most countries) and the development of Computational Social Science~\cite{lazer2009life}, enabled these `bread-crumbs' of digital traces (e.g., phone records, GPS traces, credit card transactions, webpage visits, and online social networks) to act as in situ sensors for human behavior; allowing for quantifying social actions and the study of human behavior on an unprecedented scale~\cite{eagle2006reality,gonzalez2008understanding,pentland2014social,almaatouq2016}.

Scientists have long suspected that human behavior is closely linked with socioeconomical status, as many of our daily routines are driven by activities related to maintain, to improve, or afforded by such status~\cite{becker1976economic,granovetter1973strength,granovetter1985economic}. Recent studies provided empirical support and investigated these theories in a vast and rich datasets (e.g., social media~\cite{llorente2014social}, phone records~\cite{toole2015tracking,eagle2010network}) with varying scales and granularities~\cite{blumenstock2015predicting,pappalardo2016analytical}.

In this work, we provide empirical results that support the use of Call Detail Records (CDRs) individual communication patterns to infer district-level behavioral indicators and examine their ability to explain unemployment as a socioeconomic output. In order to achieve this, we combine a large dataset of CDRs with records from the unemployment benefit program. We quantify individual behavioral indicators from over 1.8 billion logged mobile phone activities generated by 2.8 million unique phone numbers and distributed among 148 different districts in Riyadh, the capital of Saudi Arabia. We extract aggregated mobile extracted indicators (e.g., activity patterns, social interactions, and spatial markers) and examine the relationship between the district level behaviors and unemployment rates. Then, we address whether the identified variables with strong correlation suffice to explain the observed unemployment. As results, we explore the performance of several predictive models in reconstructing unemployment at the district level. Our approach is different from prior work that has already examined the relation between regional wealth and regional phone use (i.e., city~\cite{toole2015tracking,eagle2010network} or municipality~\cite{pappalardo2016analytical} level), as we focus on microregions composed of just a few households with unprecedentedly high quality ground truth labels. This type of work can provide critical input to social and economic research and policy as well as the allocation of resources.

In summary, we frame our contributions as follows:
\begin{itemize}
	\item We find that CDRs indicators are consistently associated with unemployment rates and that this relationship persists even when we include detailed controls for a district's area, population, and mobile penetration rate.
	\item We compare several categories of indicators with respect to their performance in predicting unemployment rates at the districts.
\end{itemize}

\section{Datasets}
\label{sec:Dataset}

For this study, we used an anonymized mobile phone meta data known as Call Detail Records (CDRs) and combined this with records from unemployment benefit program.

\subsection{The CDRs dataset} Consists of one full month of records for the entire country, with 3 billion mobile activities to over $10,000$ unique cell towers, provided by a single telecommunication service provider~\cite{alhasouncity,aleissawired}. Each record contains: i) an anonymized user identifier; ii) the type of activity (i.e., call or data etc); iii) the identifier of the cell tower facilitating the service; iv) duration; and v) timestamp of the activity. Each cell tower is spatially mapped to its latitude and longitude and the reception area is approximated by a/the corresponding Voronoi cell. The dataset studied records the identity of the closest tower at the time of activity; thus, we can not identify the position of a user within a Voronoi cell. For privacy considerations, user identification information has been anonymized by the telecommunication operator. Unlike standard CDRs, this dataset does not include the cell tower identity of the receiver end of the activity (i.e., only the location of the caller is approximated). The operator that provided the call data records had around 48\% market share at the time of data acquisition.

\subsection{The unemployment benefit program dataset} 
The database contains more than 4 million applications for the benefit, of which 1.4 million applications were approved, accounting for $\approx 7\%$ of the total national population. Each record contains anonymized applicant information including their home address (down to the district level). Hence, we are able to derive spatial socio-economic status of unemployed populations at the regional level (i.e., 13 Administrative areas), city level (i.e., 61 cities), and down to the district level (i.e., 1277 districts). In the present work, we focus on the 148 districts within Riyadh. 

\subsection{Census information}
Riyadh census data was obtained from the High Commission for Development of Arriyadh (ADA) at the Traffic Analysis Zones (TAZs) level. The administrative areas and city level census information were matched using their identifier codes. The district level information was obtained by mapping the TAZ information to the district boundaries. The average spatial resolution (i.e., square root of the land area divided by the number of land units) for the districts and TAZs in Riyadh is $2.6 km$ and $0.04 km$, respectively.

\subsection{Mapping Census Population to Districts}
For the $i^{th}$ TAZ denoted by $\tau_i$ we have the population $P_{\tau_i}$ and demographic breakdown (i.e., gender and nationality), as well as housing data (i.e., number of houses, villas, apartments etc.) provided by the ADA. However, the finest resolution for the unemployment data is at the district level. Therefore, for each district $d_i$ we estimate the population $P_{d_i}$ as follows:

$$
P_{d_i} = \sum_{j=0}^{|\tau|} \frac{A_{(d_i \cap \tau_j)} P_{\tau_j}} {A_{\tau_j}}
$$

where $|\tau|$ is the total number of TAZ units, $A_{\tau_j}$ is the area of the $j^{th}$ TAZ unit and $A_{(d_i \cap \tau_j)}$ is the intersection area of $d_i$ and $\tau_j$.

\subsection{Mapping Mobile Population to Districts}
For each cell tower $c_j$, we know the total number of different users $T_{c_j}$ with home location (i.e., the tower where a user spends most of the time at night; as in~\cite{Santi2012}) being the $j^{th}$ tower. When one makes a phone call, the network usually identifies nearby towers and connects to the closest one. The coverage area of a tower $c_j$ thus was approximated using a Voronoi-like tessellation. The Voronoi cell associated with tower $c_j$ is denoted by $v_j$. Therefore, we can compute the penetration rate $\sigma_{d_i}$ for district $i$ as follows:

$$
\sigma_{d_i} = \frac{1}{P_{d_i}} \sum_{j=0}^{|v|} \frac{A_{(d_i \cap v_j)} T_{c_j}} {A_{v_j}}
$$

where $|v|$ is the total number of Voronoi cells, $A_{v_j}$ is the area of the $j^{th}$ Voronoi cell (associated with the $j^{th}$ cell tower) and $A_{(d_i \cap v_j)}$ is the intersection area of $d_i$ and $v_j$.

\begin{figure}[h!]
	\centering
	\includegraphics[width=1\columnwidth]{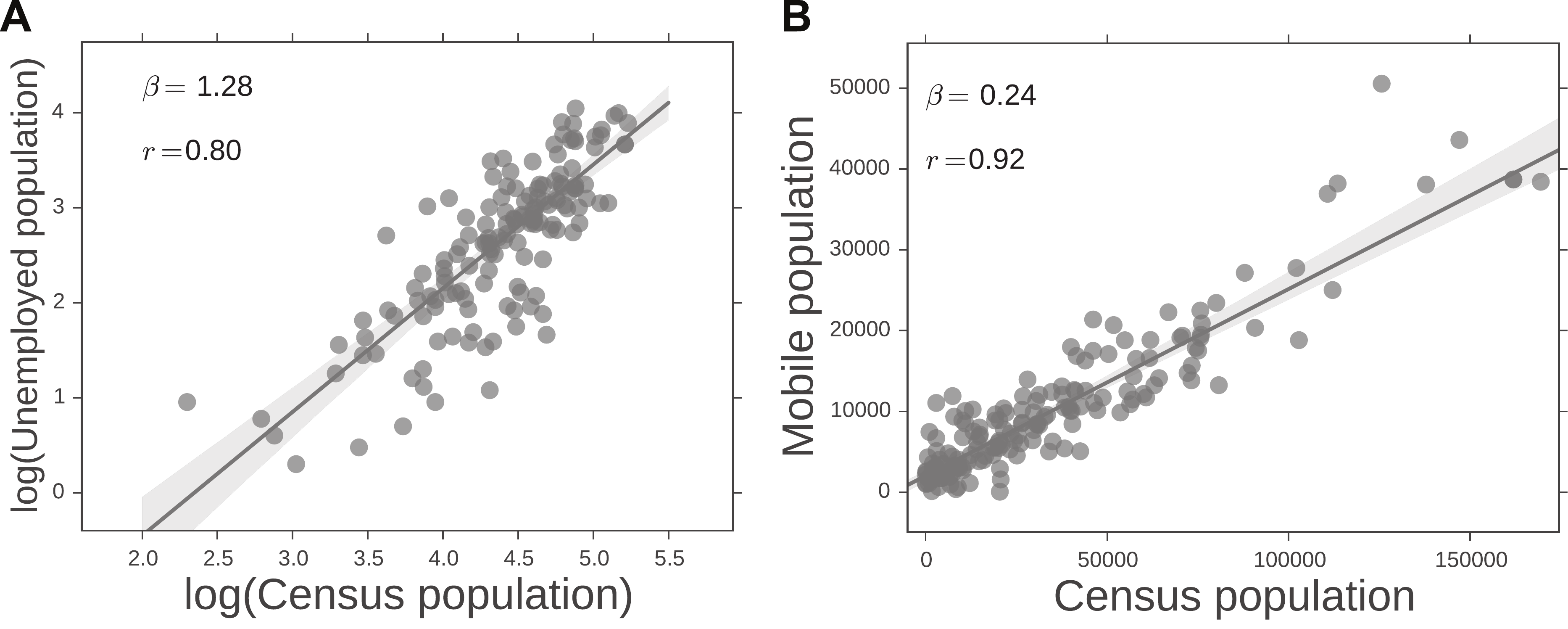}
	\caption{Panel \textbf{(A)} shows the relationship between unemployment and population in 2012, for the 148 districts. Panel \textbf{(B)} shows the number of homes detected, for the 148 districts versus district census population. Best-fit scaling relations are shown as solid lines and the 95\% confidence intervals are shown as a shaded area.}
	\label{fig:mapping_population}
\end{figure}

Figure~\ref{fig:mapping_population} shows the scaling relationships between the district population versus unemployment rate and also population versus mobile users. These results are consistent with previous studies indicating that scaling with population is indeed a pervasive property of urban organization~\cite{bettencourt2007growth,pan2013urban}.

\section{Extracting Behavioral Indicators}
\label{sec:indicators}
The goal of this work is to investigate how behavioral indicators from mobile phone meta data can be extracted and then related back to the economical wellbeing of geographical regions (i.e., districts). To this end, we define three groups of indicators that have been widely explored in fields like economy or social sciences. Several of these indicators have been implemented in the bandicoot toolbox~\cite{de2013predicting}\footnote{Bandicoot can be found at \url{http://bandicoot. mit.edu/docs/}}. All the indicators are computed at the individual level and then aggregated and standardized (i.e., scaled to have mean zero and variance one) at the district level.

\subsection{Activity patterns}

Activity patterns quantify factors related to the aggregate patterns of mobile usage for each district such as volume (average number of records per user), timing (the average percentage of night calls -- nights are 7pm-7am), and duration (average duration of calls).

\subsection{Social interactions}

The social interaction indicators capture the structure of the individual's contact network. We focus on the egocentric networks around the individual in order to examine the local structure and signify the types of interactions that develop within their circle~\cite{almaatouq2014twitter}. 

Let $E_i = (\mathcal{C}_i,\mathcal{E}_i)$ be the directed egocentric-graph that represents the topological structure of the $i^{th}$ individual where $\mathcal{C}_i$ is the set of contacts (total number of contacts is $k = \vert \mathcal{C}_i\vert $) and $\mathcal{E}_i$ is the set of edges. A directed edge is an ordered pair $(i,j)$ with associated call volume $w_{ij}$ (and/or $(j,i)$  with $w_{ji}$ volume) representing the interaction between the ego $i$ and a contact $ j \in \mathcal{C}_i$. Note that by definition $w_{ij} + w_{ji} \in \mathbb{Z}^+$, must be satisfied, otherwise $j \notin \mathcal{C}_i$. Therefore, the volume $w_{ij}$ is set to $0$ when $(i,j) \notin \mathcal{E}_i$, alternatively $w_{ji} = 0$, if the $(i,i)$ direction does not exist. From this we can compute several indicators for an individual within its egocentric network. We define $I(i)$ and $O(i)$ as the set of incoming interactions to (respectively, initiating from) individual $i$. That is,  
$$I(i) = \{j \in \mathcal{C}_i \vert (j,i) \in \mathcal{E}_i \}, \quad \text{and} \quad  O(i) = \{j \in \mathcal{C}_i \vert (i,j) \in \mathcal{E}_i \} \text{.}$$

\paragraph{Percentage of initiated interaction} is a measure of directionality of communication. We define it as $$\mathcal{I}(i) = \vert I(i)\vert / \left(\vert I(i)\vert +\vert O(i)\vert\right)$$ 

\paragraph{Balance of contacts} is measured through the balance of interactions per contact. For an individual $i$, the balance of interactions is the number of outgoing interactions divided by the total number of interactions.

\begin{equation*}
\beta(i) = \frac{1}{k} \sum_{j \in I(i) \cup O(i)} \frac{w_{ij}}{w_{ji} + w_{ij}}
\end{equation*}

\paragraph{Social Entropy} captures the social diversity of communication ties within an individual's social network, we follow Eagle's et al.~\cite{eagle2010network}   approach by defining social entropy, $D_{social}(i)$, as the normalized Shannon entropy associated with the $i^{th}$ individual communication behavior:

\begin{equation*}
D_{social}(i) = -\sum_{j\in \mathcal{C}_i}log(p_{ij})/{log(k)}
\end{equation*}

Where $p_{ij} = w_{ij}/{\sum_{\ell=1}^k w_{i\ell}}$ is the
proportion of $i$'s total call volume that involves $j$. High diversity scores imply that an individual splits his/her time more evenly among social ties.

\subsection{Spatial markers}
The spatial markers captures mobility patterns and migration based on geospatial markers in the data. In this work, we measure the number of visited locations, which captures the frequency of return to previously visited locations over time~\cite{song2010modelling} (time in our case is the entire observational period). We also compute the percentage of time the user was found at home.

\subsection{Unsupervised Clustering}
\label{som}
We use the standard form of self-organizing maps (SOMs) as an unsupervised clustering analysis tool~\cite{kohonen1998self,wehrens2007self}.

\begin{figure}[h!]
	\centering
	\includegraphics[width=1\columnwidth]{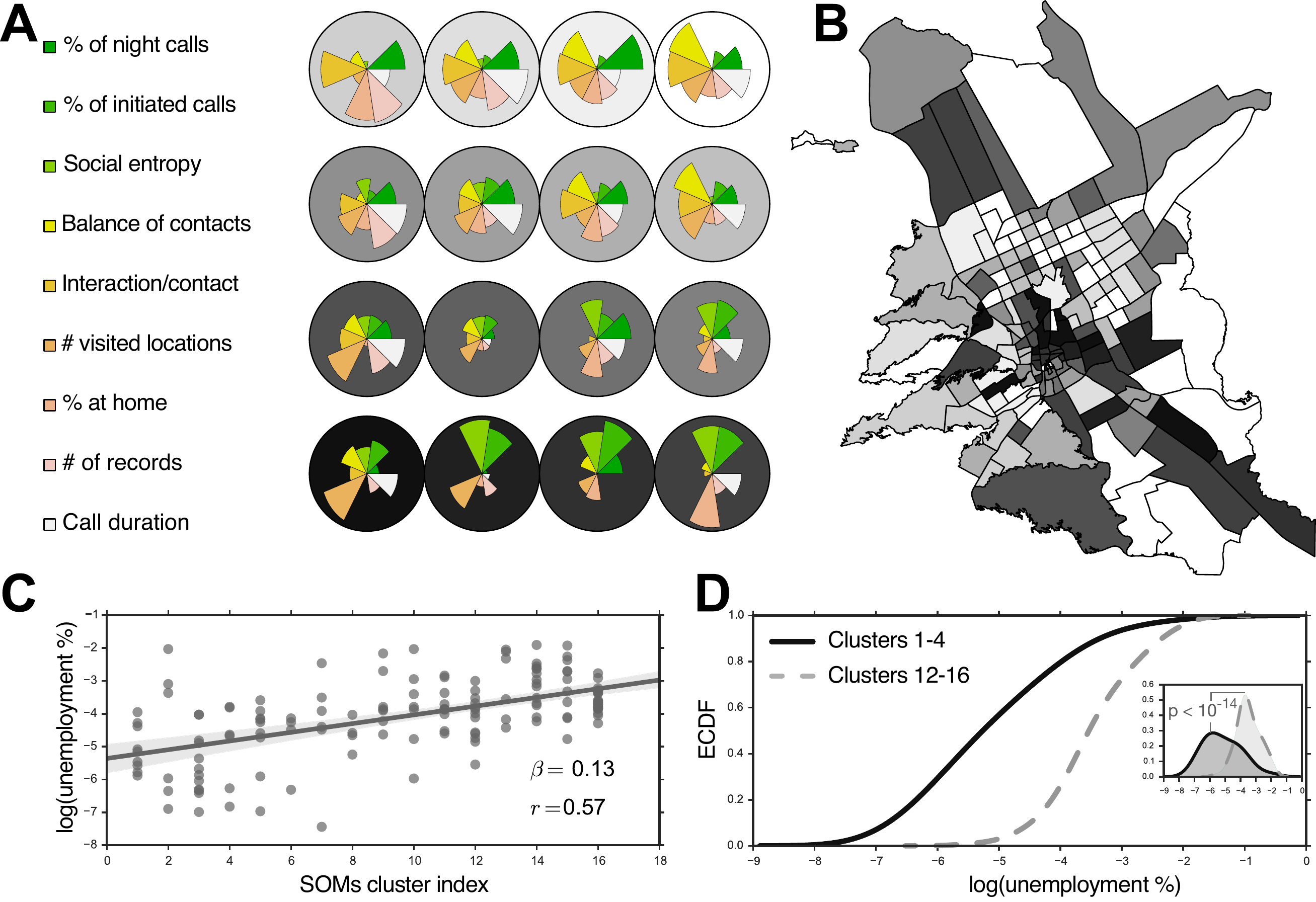}
	\caption{Panel~\textbf{(A)} shows the plot of the color-encoded codebook vectors of the 4-by-4 mapping of the districts behaviors. Panel~\textbf{(B)} shows the combined view of attribute space (i.e., SOMs results) and geographic space (choropleth map). Panel \textbf{(C)} demonstrates the relationship between the clustered behaviors and unemployment rates. Finally, panel \textbf{(D)} shows that the Empirical Cumulative Distribution Function (ECDF) of the unemployment rates for contrasting groups behaviors.}
	\label{fig:som}
\end{figure}

In Figure~\ref{fig:som}A, the codebook vectors from the resulting SOMs are shown in a segments plot, where the grayscale background color of a cluster corresponds to its index (i.e., number of clusters = 9 arranged in a rectangular grid). Districts having similar characteristics based on the multivariate behavioral attributes are positioned close to each other, and the distance between them represents the degree of behavioral similarity or dissimilarity. High average spatial entropy with small percentage of time being at home, for example, is associated with districts projected in the bottom left corner of the map (i.e., cluster index one -- black color).  On the other hand, districts with low social entropy, percentage of initiated calls, and balance of contacts are associated with the clusters at the top column of the map. On the geographic map (see Figure~\ref{fig:som}B), each district is assigned a color, where the meaning of the color can be interpreted from the corresponding codebook vector. 
We can see that at the center of the city, most districts are assigned to clusters with dark backgrounds, and the color gets lighter as we move towards the periphery of the city. As expected from the description of the corresponding codebook vectors, districts projected in the bottom of the map (dark color background) are associated with lower unemployment rates (see Figure~\ref{fig:som}C). It is indeed the case that districts with similar behavioral attributes have similar unemployment rates (see Figure~\ref{fig:som}.D).

\subsection{Statistical Correlation}
\label{sec:correlation}

\begin{figure}[h!]
	\centering
	\includegraphics[width=1\columnwidth]{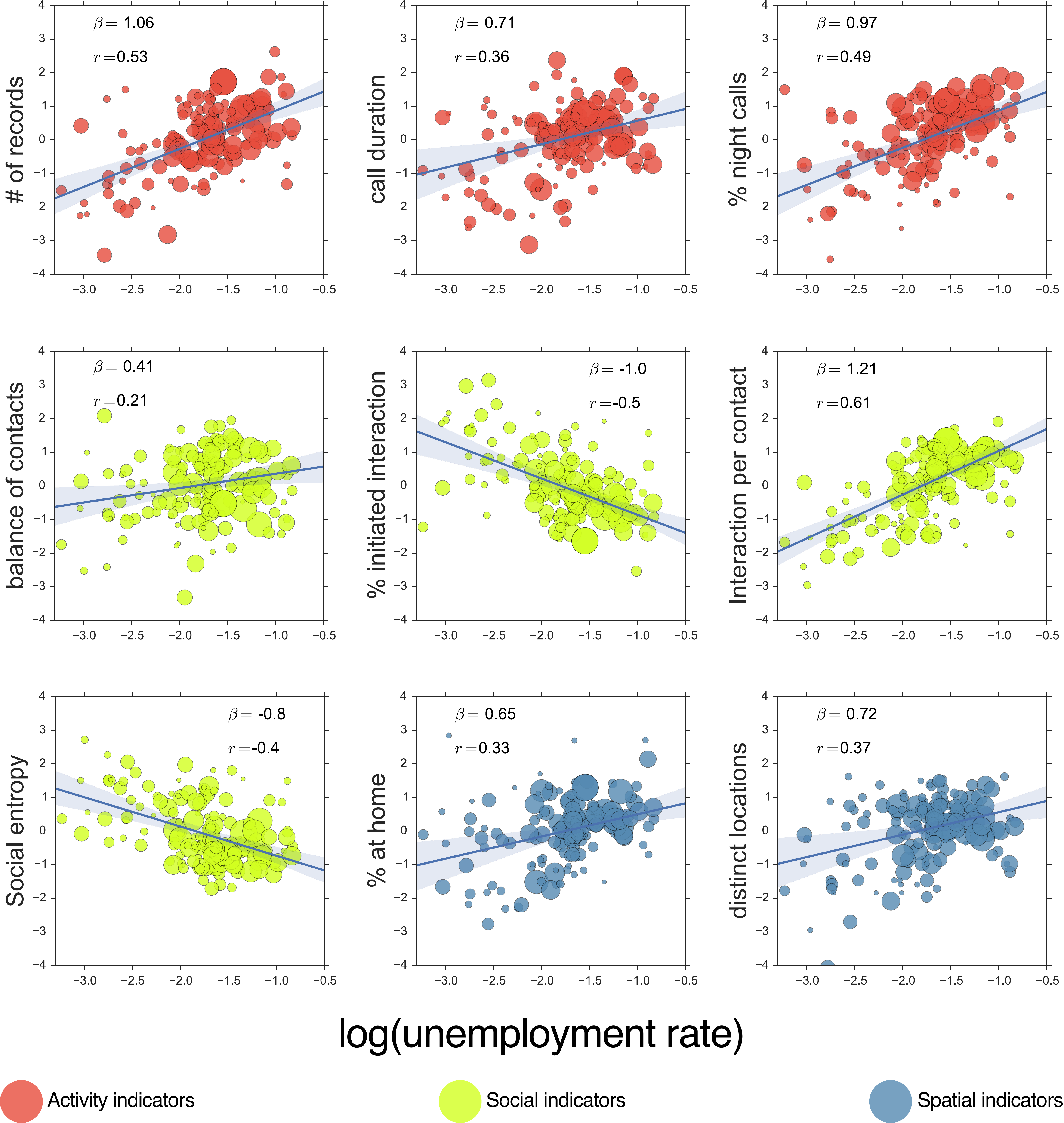}
	\caption{Relations between the mobile extracted indicators (district average) for the 148 districts against its unemployment rate. Size of the
		points is proportional to the population in each district. Solid lines correspond to linear fits to the data and the shaded area represents the 95\% confidence intervals.}
	\label{fig:correlations}
\end{figure}

As we can see in Figure~\ref{fig:correlations}, all the extracted indicators exhibit at least moderate statistical correlations with unemployment. In addition, we find that the indicators relationship with unemployment persists for most indicators even when we include controls for a district's area, population, and mobile penetration rate (see Table~\ref{tab:4}). These results suggest that several of those indicators are sufficient to explain the observed unemployment.
\begin{table}[h!]
	\begin{center}
		\scriptsize
		\begin{tabular}{l D{.}{.}{3.5}@{} D{.}{.}{3.4}@{} D{.}{.}{3.5}@{} D{.}{.}{3.5}@{} D{.}{.}{3.5}@{} D{.}{.}{3.4}@{} D{.}{.}{3.5}@{} D{.}{.}{3.4}@{} D{.}{.}{3.4}@{} }
			\toprule
			& \multicolumn{9}{c}{Dependent variable: log(Unemployment Rate)}\\
			\toprule
			& \multicolumn{1}{c}{(1)} & \multicolumn{1}{c}{(2)} & \multicolumn{1}{c}{ (3)} & \multicolumn{1}{c}{(4)} & \multicolumn{1}{c}{(5)} & \multicolumn{1}{c}{(6)} & \multicolumn{1}{c}{(7)} & \multicolumn{1}{c}{(8)} & \multicolumn{1}{c}{(9)} \\

			\midrule
			Population                & 0.15       & 0.16      & 0.12        & -0.07      & 0.09       & 0.12       & 0.12       & 0.16      & 0.16      \\
			& (0.10)     & (0.11)    & (0.11)      & (0.11)     & (0.11)     & (0.11)     & (0.11)     & (0.11)    & (0.11)    \\
			Area                      & 0.07       & 0.15^{*}  & 0.12        & 0.28^{***} & 0.14^{*}   & 0.17^{*}   & 0.11       & 0.20^{**} & 0.19^{**} \\
			& (0.09)     & (0.09)    & (0.08)      & (0.08)     & (0.08)     & (0.08)     & (0.09)     & (0.08)    & (0.08)    \\
			Penetration rate & 0.02       & 0.00      & 0.04        & 0.04       & -0.06      & 0.02       & 0.05       & -0.02     & 0.02      \\
			& (0.11)     & (0.11)    & (0.11)      & (0.10)     & (0.10)     & (0.11)     & (0.11)     & (0.11)    & (0.11)    \\
			\# of records              & 0.31^{***} &           &             &            &            &            &            &           &           \\
			& (0.09)     &           &             &            &            &            &            &           &           \\
			Call duration             &            & 0.17^{**} &             &            &            &            &            &           &           \\
			&            & (0.08)    &             &            &            &            &            &           &           \\
			\% initiated inter.  &            &           & -0.28^{***} &            &            &            &            &           &           \\
			&            &           & (0.09)      &            &            &            &            &           &           \\
			\% night calls            &            &           &             & 0.49^{***} &            &            &            &           &           \\
			&            &           &             & (0.09)     &            &            &            &           &           \\
			\% at home                &            &           &             &            & 0.30^{***} &            &            &           &           \\
			&            &           &             &            & (0.08)     &            &            &           &           \\
			Social entropy            &            &           &             &            &            & -0.18^{**} &            &           &           \\
			&            &           &             &            &            & (0.09)     &            &           &           \\
			Inter. per contact   &            &           &             &            &            &            & 0.27^{***} &           &           \\
			&            &           &             &            &            &            & (0.09)     &           &           \\
			Balance of contacts       &            &           &             &            &            &            &            & -0.01     &           \\
			&            &           &             &            &            &            &            & (0.08)    &           \\
			visited locations        &            &           &             &            &            &            &            &           & 0.14^{*}  \\
			&            &           &             &            &            &            &            &           & (0.08)    \\
			(Intercept)                 & -0.00      & 0.00      & -0.00       & 0.00       & -0.00      & -0.00      & -0.00      & -0.00     & -0.00     \\
			& (0.08)     & (0.08)    & (0.08)      & (0.07)     & (0.08)     & (0.08)     & (0.08)     & (0.08)    & (0.08)    \\
			\midrule
			R$^2$                     & 0.15       & 0.10      & 0.14        & 0.23       & 0.16       & 0.10       & 0.13       & 0.07      & 0.09      \\
			Adj. R$^2$                & 0.13       & 0.08      & 0.12        & 0.21       & 0.13       & 0.07       & 0.11       & 0.05      & 0.07      \\
			BIC                       & 421.97     & 430.53    & 424.03      & 407.25     & 420.75     & 430.65     & 425.16     & 434.73    & 431.86    \\
			Num. obs.                 & 147        & 147       & 147         & 147        & 147        & 147        & 147        & 147       & 147       \\
			\bottomrule
			\multicolumn{10}{l}{\scriptsize{$^{***}p<0.01$, $^{**}p<0.05$, $^*p<0.1$. Standard errors in parentheses.}}
		\end{tabular}
		\vspace{2mm}
		\caption{Regression table explaining the districts' unemployment rate as a function of the activity patterns, social interactions, and spatial markers, with the inclusions of controls for a district's area, population, and mobile penetration rate.}
		\label{tab:4}
	\end{center}
\end{table}
For instance, we find the percentage of night calls to have the highest effect size and explanatory power ($R^2 = 23\%$; model 4). This is expected, as regions with very different unemployment patterns should exhibit different temporal activities. Since working activities usually happen during the day, we would expect that districts with high levels of unemployment will tend to have higher proportion of their activities during the night.

Previous study~\cite{schneider2013unravelling} have found that the duration spent at either home or work is relatively flat distributed with peaks around time spans of 14 hours at home and 3.5-8.6 hours at work. Therefore, we hypothesize that the lack of having a work location for the unemployed would lead to an increase in the duration spent at home (i.e., \% home), and/or reduce the tendency for revisiting locations (i.e., higher visited locations). We indeed find that the percentage of being home and number of visited locations to be associated with unemployment in our dataset.

We also find the percentage of initiated interactions to be negatively correlated with unemployment. This indicators has been shown to be predictive of the Openness (i.e, the tendency to be intellectually curious, creative, and open to feelings) personality trait~\cite{john1999big,de2013predicting}, which in return is predictive of success in job interviews~\cite{caldwell1998personality}.

As in~\cite{eagle2010network,llorente2014social}, we find that districts with high unemployment rates have less diverse communication patterns than areas with low unemployment. This translates in a  negative coefficient for social entropy and positive coefficient for the interaction per contact indicator. The balance of contacts factor was not found to be significant ($p>0.1$).

\subsection{Supervised Predictive Model}

Here we are interested in the predictability of unemployment rates of microregions based on the mobile phone extracted indicators and independently of additional census information such as population, gender, income distribution, etc.
Such additional information is often  unavailable in developing nations, which by itself represents a major challenge to policy-makers and researchers. Therefore, it is of utmost importance to find novel sources of data that enables new approaches to demographic profiling.

We analyze the predictive power of the indicators using Gaussian Processes (GP) to predict unemployment based on mobile phone indicators solely. We train and test the model in K-fold-cross validation ($K=5$) and compute the coefficient of determination $R^2$ as a measure of quality for each category of indicators (i.e., activity, social, and spatial) and also for the full indicators (involving all mobile extracted indicators presented in this work). The advantage for using Gaussian Processes (GP) to regress unemployment rates is that the model produces probabilistic (Gaussian) predictions so that one can compute empirical confidence intervals and probabilities that might be used to refit (online fitting, adaptive fitting) the prediction in some region of interest~\cite{welch1992screening,almaatouq2016complex,nielsen2002dace}.

\begin{figure}[h!]
	\centering
	\includegraphics[width=1\columnwidth]{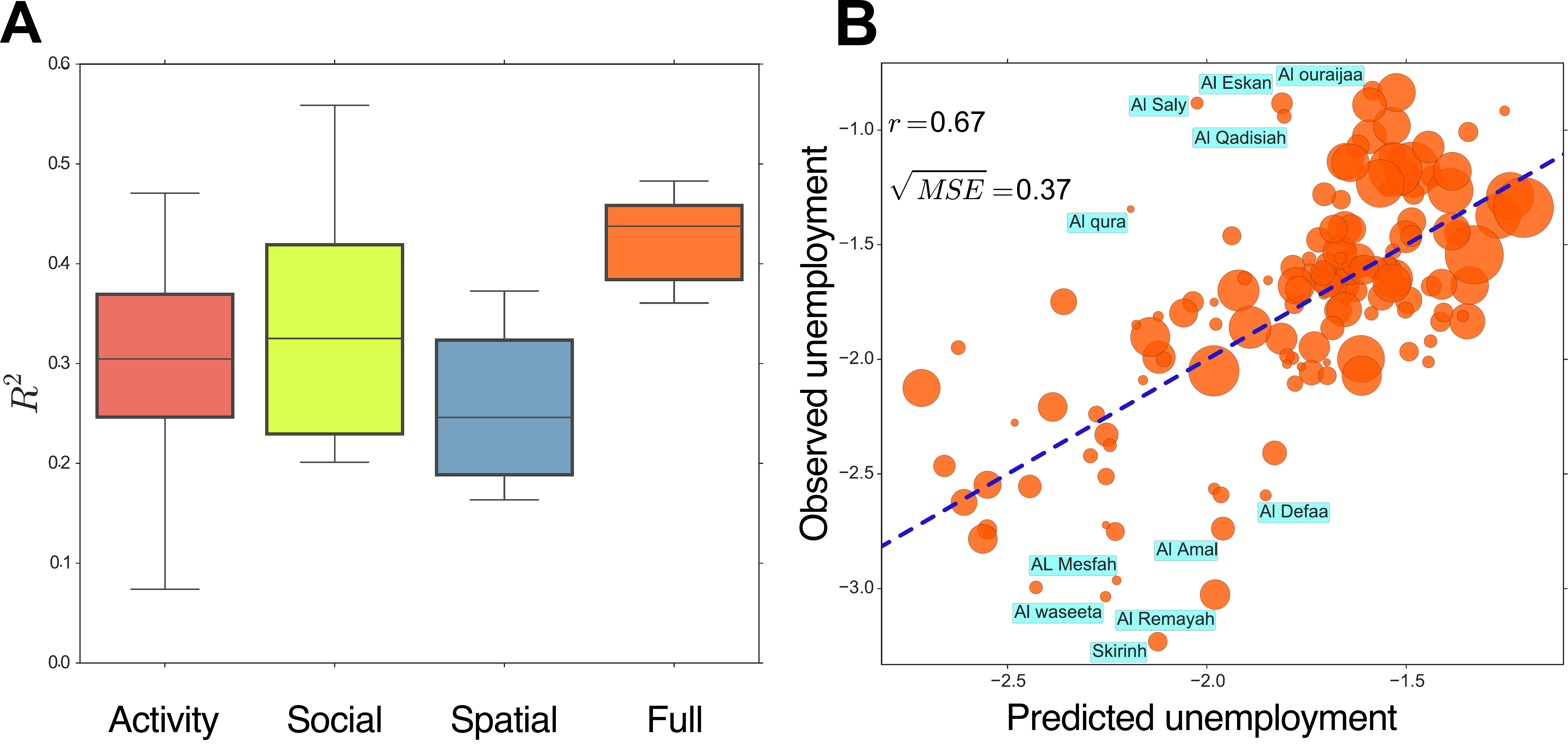}
	\caption{Panel \textbf{(A)} shows the average performance of each indicator category in predicting unemployment. Panel \textbf{(B)} depicts the cross-validated prediction of unemployment rates versus the observed ones, $r= 0.68$. The predicted values are based on the prediction that was obtained for that district when it was in the test set. Dashed line correspond to the equality line.}
	\label{fig:prediction}
\end{figure}

In Figure~\ref{fig:prediction}A we find that the social interaction indicators to be very predictive of unemployment with an average $R^2 = 0.43$ (95\% CI: 0.37 -- 0.48), which is more predictive than the activity pattern indicators $R^2 = 0.29$ (95\% CI: 0.15 -- 0.39) and the spatial indicators $R^2 = 0.26$ (95\% CI:0.19 -- 0.33). It is worth mentioning, that the composite model performed significantly better than single category models with an average $R^2= 0.43$ (95\% CI: 0.37 -- 0.48). Figure~\ref{fig:prediction}.B compares the predicted and observed unemployment rate for each based on the prediction that was obtained for that district when it was in the test set.

\section{Summary \& Future Work}
\label{sec:conclusion}

In this paper we have demonstrated that mobile phone indicators are associated with unemployment rates and that this relationship is robust to the inclusion of controls for a district's area, population, and mobile penetration rate. Following this analysis, we also investigated the predictability of unemployment rates with respect to three categories of indicators, namely, activity patterns, social interactions, and spatial markers. The results of these analyses highlighted the importance of social interaction indicators for predicting unemployment. 

Note that we are not stating a causality arrow between the indicators and the unemployment rate as we do not have individual level mapping of unemployment with which to test for individual differences. In this work, our goal is to show that aggregate behavioral indicators of the members of a district represent a strong statistical signature that can be used as alternative measuring approach with a real translation in the economy.

In our future work, we intend to intersect the Call Detail Records (CDR) and unemployment data derived from the unemployment benefit program at the individual level. This will allow for the study of how the behavioral signature of a single individual can be used to predict that same individual's employment status. This could reveal the key determinants of unemployed people to find a job and allow for designing personalized intervention mechanisms.





\section*{Acknowledgments}
The authors thank the Center for Complex Engineering Systems (CCES) at KACST and MIT and the Media Lab at MIT for their support.

\bibliographystyle{splncs03}

\end{document}